\documentclass[aps,prl,twocolumn,showpacs,floatfix]{revtex4}
\usepackage{amsmath}
\usepackage{graphicx,epsfig,psfrag}
\usepackage{amssymb}
\begin{document}

\title{Beyond Planck-Einstein quanta: amplitude driven quantum excitation}
\author{Wen Shen$^1$}
\author{T. P. Devereaux$^{2,3}$}
\author{J. K. Freericks$^{1,4}$}
\affiliation{
$^{1}$ Department of Physics, Georgetown University, Washington D.C., 20057, USA\\
$^2$SLAC National Accelerator Laboratory, Stanford University, Stanford, CA 94305, USA\\
$^3$Geballe Laboratory for Advanced Materials, Stanford University, Stanford, CA 94305, USA\\
$^4$Kavli Institute for Theoretical Physics, Santa Barbara, CA 93106, USA
}
\begin{abstract}
Linear-response quantum excitation is proportional to the amplitude of the field, with the energy of the excitation given by the driving frequency. As the amplitude is increased, there is a crossover, where the excitation energy is governed by the amplitude of the driving field, not its frequency.  As the amplitude is increased even further, then complex quantum oscillations develop, whose origin is related to Wannier-Stark physics, but has no simple explanation. We illustrate this phenomena with the exact solution of the simplest model of a charge-density-wave insulator driven by a uniform time-dependent electric field.
\end{abstract}

\maketitle

\paragraph{Introduction}
In 1901, Planck introduced the idea of light quanta to calculate the spectrum for black body radiation~\cite{planck}, which was employed by Einstein in 1905 to explain the mysterious quantum properties of the photoelectric effect~\cite{einstein}. Later, the solution of the Landau-Zener tunneling problem in the 1930s~\cite{landau,zener}, where a tunneling excitation is determined by the speed at which the minimal excitation gap is approached (and is proportional to the amplitude of an effective driving field) showed how the amplitude and not the frequency of a driving field can govern quantum excitation beyond linear response. 

We consider one of the simplest problems in driven quantum systems: the excitation of an insulator across its gap due to a monochromatic applied ac electric field. The energy associated with this driving frequency satisfies the Planck-Einstein relation~\cite{planck,einstein}, $E=\hbar\omega$, and is independent of the amplitude of the driving field, hence we expect to see no response until the frequency is large enough that $\hbar\omega\ge E_{\rm gap}$. Linear response theory verifies this result, as the Kubo-Greenwood formula shows that the amplitude of the field just provides an overall scale to the response~\cite{kubo,greenwood}, and the ability to create an excitation is determined by energy conservation.  As the amplitude of the field is increased, photons of a lower energy can combine together and create higher energy photons, hence one would expect resonances at $\hbar\omega/2$ (or generally $\hbar\omega/n$ for multiphoton processes). However, the presence of a large field can also modify the quantum states themselves, and create states inside the gap region, thereby reducing the effective  gap, and allowing excitations to occur at even lower frequencies. 

The Landau-Zener tunneling problem has investigated some of these aspects~\cite{landau,zener}. While there is no applied field {\it per se} in this problem, one can assume that the rate at which the gap is approached is proportional to an effective driving field, and in this situation, it is known that the efficiency in tunneling to create excitations across the gap depends exponentially on the driving rate, and hence on the amplitude of the effective field. Indeed, the adiabatic theorem guarantees no excitation for infinitesimally small fields with $\omega\rightarrow 0$. Alternatively, if one thinks about the problem semiclassically, then when the energy gain from the amplitude of the electric field over one lattice spacing is equal to the gap energy, one expects the electron to be able to transfer across the gap from purely classical energetics, as the transition now becomes allowed. 

The quantum excitation problem has been examined in the context of field-induced ionization of atoms, and in the photoelectric effect.  For the atomic problem, Keldysh~\cite{keldysh} showed how a detailed quantum theory can describe the full regime from the frequency-driven excitation to the amplitude-driven excitation, and these ideas have been applied recently to experiments on gold nanotips in the photoelectric effect, where multi-photon excitations are observed when the photon light field has a large amplitude due to enhancement near an isolated sharp metallic tip~\cite{experiment}. But the quantum excitation problem in solid state systems is more complex, because there are significant final-state effects due to the Pauli exclusion principle. As an electron is excited from a lower band to an upper band, the state the particle has been excited to is blocked from further excitation.  Such effects become quite important as the field becomes strong enough to excite significant fractions of electrons from the lower to the upper band. In addition, Bragg scattering in the crystal gives rise to Bloch oscillations, which can both excite and de-excite electrons across the gap. These effects greatly complicate the net quantum excitation process. 

Two questions immediately come to mind about field-driven quantum excitations in this context: (i) does the excitation continue to depend exponentially on the amplitude so that nearly all of the excitation occurs near the maximal amplitude of a pulsed field and (ii) as the amplitude is increased do we find a regime of purely amplitude-driven tunneling, where the excitation becomes independent of the driving frequency? We answer both of these questions with an exact nonequilibrium solution for the nonlinear excitation of a quantum solid. Note that this model has no interactions and does not thermalize.
If the gap is large enough, the populations in the upper and lower bands will hardly change, and the only effect of thermalization will be to redistribute the occupancy of the energy levels in each band. This process, which is ignored here, is a longer-time process that is unlikely to significantly modify the phenomena that is shown here.

\begin{figure}[!htb]
\begin{center}
\includegraphics[scale=0.20]{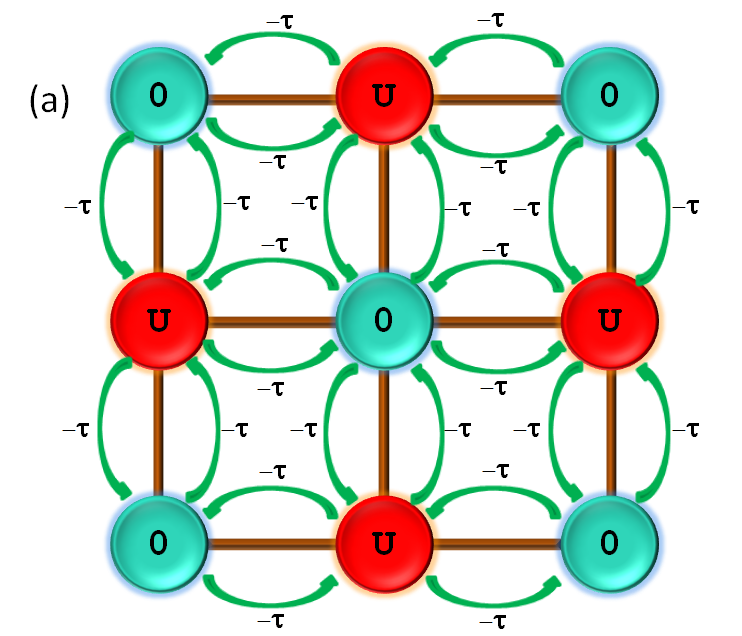}
\includegraphics[scale=0.25,clip=on]{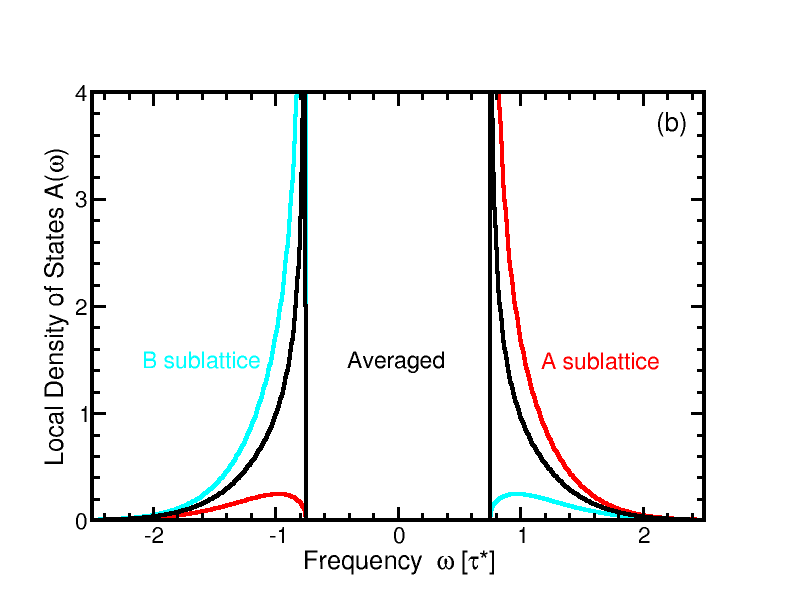}
\end{center}
\caption{(Color online) 
(a) Schematic of the charge-density-wave model on the checkerboard lattice.  The A sublattice (red) has the on-site potential $U$, while the B sublattice (light blue) has a vanishing potential. Hopping is between nearest neighbors as indicated by the green arrows. The schematic shows a two-dimensional square lattice, but we work on an infinite-dimensional hypercubic lattice. (b) Local density of states on the A/B sublattices (red/light blue) and the average density of states (black) in equilibrium for $U=1.5$.
\label{fig: lattice}}
\end{figure}

\paragraph{Formalism}
The model we employ for the CDW is that of spinless electrons moving on a lattice with a bipartite structure (a checkerboard) that has different site energies on the two sublattices ($A/B$). The Hamiltonian in the Schr\"odinger picture is
\begin{equation}
\mathcal{H}(t)=-\sum_{ij}\tau_{ij}(t)c_i^\dagger c_j^{}+\sum_{i\in A}(U-\mu)c_i^\dagger c_i^{}+\sum_{i\in B}(-\mu)c_i^\dagger c_i^{}.
\label{eq: ham}
\end{equation}
The first term is the kinetic energy, which involves a hopping between nearest neighbor lattice sites $i$ and $j$ with a hopping integral $\tau_{ij}(t)$ (the hopping matrix is a Hermitian matrix that is nonzero only for $i$ and $j$ nearest neighbors).  The second and third terms include the chemical potential $\mu$ and the external potential $U$ which is nonzero only on the $A$ sublattice. We set $\mu=U/2$ in our calculations to have the case of half filling.  Since the electrons do not interact with each other, the spin degree of freedom is trivial, and has been neglected here. The field is introduced via a time-dependent hopping integral, which becomes time-dependent due to the Peierls’ substitution~\cite{peierls}. The hopping integral then is described by the following time-dependent function:
\begin{eqnarray}
\tau_{ij}(t)&=&\frac{\tau^*}{2\sqrt{d}}\exp\left [ -\frac{ie}{\hbar c}\int_{{\bf R}_i}^{{\bf R}_j}{\bf A}(t)\cdot d{\bf r}\right ]\nonumber\\
&=&
\frac{\tau^*}{2\sqrt{d}}\exp\left [\frac{ie}{\hbar c}({\bf R}_i-{\bf R}_j)\cdot {\bf A}(t)\right ].
\label{eq: hopping}
\end{eqnarray}
and we take the limit as $d\rightarrow\infty$ using $\tau^*$ as the energy unit. Here ${\bf A}(t)$ is the time dependent (but spatially uniform) vector potential in the Hamiltonian gauge (where the scalar potential vanishes). The field is chosen to point in the diagonal direction ${\bf A}(t) = A_0(t)(1, 1, 1 \ldots )$ with $A_0(t)$ given by the antiderivative of the electric field as a function of time. In this case, the (time-dependent) electronic band structure becomes
\begin{equation}
\epsilon\left ( {\bf k}-\frac{e{\bf A}(t)}{\hbar c}\right )=\lim_{d\rightarrow\infty}\frac{\tau^*}{\sqrt{d}}\sum_{i=1}^d\cos\left [ k_i-\frac{e{\bf A}_i(t)}{\hbar c}\right ],
\label{eq: peierls}
\end{equation}
which is the standard form for the Peierls substitution~\cite{peierls}. 

\begin{figure}[!ht]
\begin{center}
\includegraphics[scale=0.25,clip]{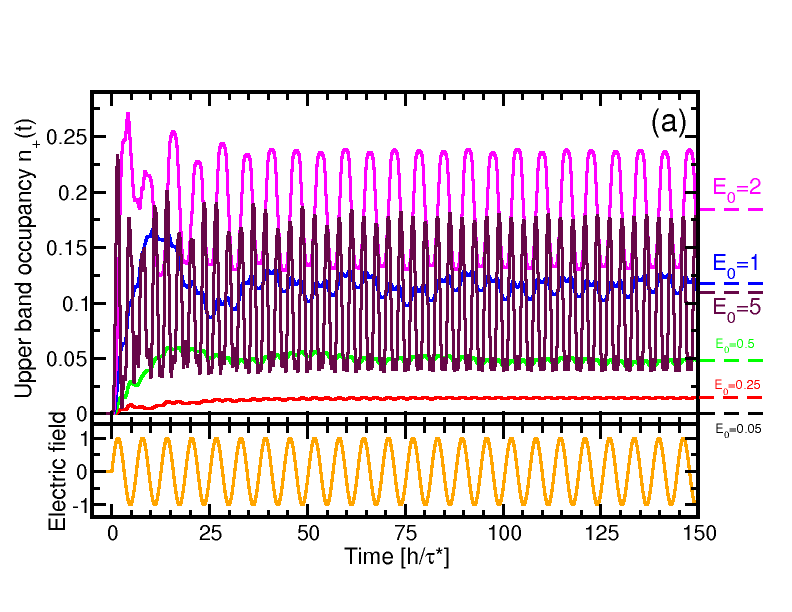}
\vskip -0.14in
\includegraphics[scale=0.25,clip]{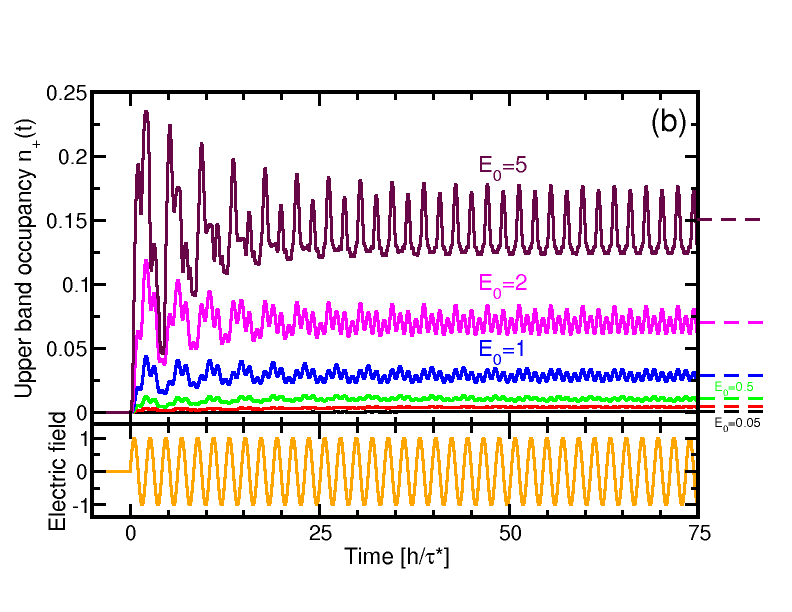}
\end{center}
\caption{(Color online) 
Time trace for the occupancy in the upper band as a function of time for an ac field driving of the CDW insulator with an equilibrium gap equal to $U=1.5$ (started initially from $T=0$) and two frequencies: (a)  $\omega_0=1$ (less than the equilibrium gap) and (b) $\omega_0=3$ (larger than the equilibrium gap).  We take as our measure of the final occupancy of the upper band the average value of the occupancy, averaged over one period in the long time limit.  These values are indicated by the dashed lines. The labels show the amplitude of the field.  The red curve in panel (b) is for $E_0=0.25$. The field trace is plotted below each panel.  Note that the time axis is half as long in panel (b).
\label{fig: ac_excite}}
\end{figure}

 For concreteness, we work in the infinite-dimensional limit, although the procedure produces an exact solution in any dimension (since the only effect of dimensionality is on the shape of the normal state density of states, we expect the infinite-dimensional results to be very similar to those in two or higher dimensions; we choose this limit because it will make for easier comparison with dynamical mean-field calculations in the ordered phase). We take $\tau=\tau^*/2\sqrt{d}$ and set $\tau^*=1$ as the energy unit). The bandstructure in the absence of a field has a gap of size $E_{\rm gap}=U$, with a density of states that is a mirror image on the $A$ and $B$ sublattices (see Fig.~\ref{fig: lattice}). There is a square-root-like singularity at the upper or lower band edge for the local density of states on each sublattice. The applied field is chosen to be spatially uniform but time dependent, with either a monochromatic ac field turned on at time $t=0$ or an oscillating pulse shape. The problem can be solved exactly by employing the Kadanoff-Baym-Keldysh formalism for the contour-ordered Green’s function and using the Trotter formula to evaluate the relevant $2\times 2$ evolution operators for each coupled momenta {\bf k} and ${\bf  k}+{\bf Q}$ with ${\bf  Q}=(\pi,\pi,\pi,\ldots)$, as described in detail in Ref.~\cite{cdw_long}. 

\begin{figure}[!ht]
\begin{center}
\includegraphics[scale=0.31,clip=on]{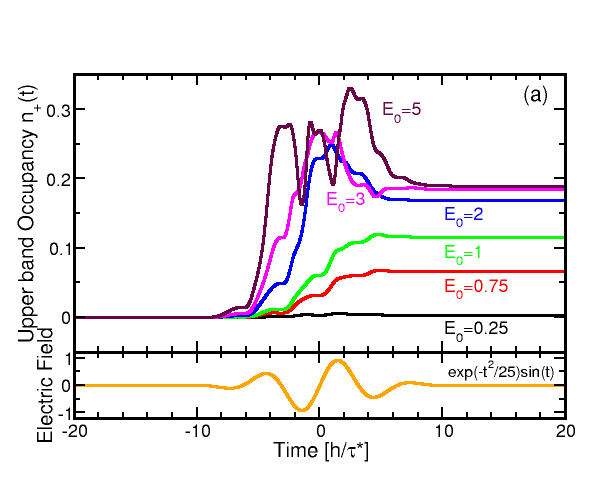}
\vskip -0.14in
\includegraphics[scale=0.25,clip=on]{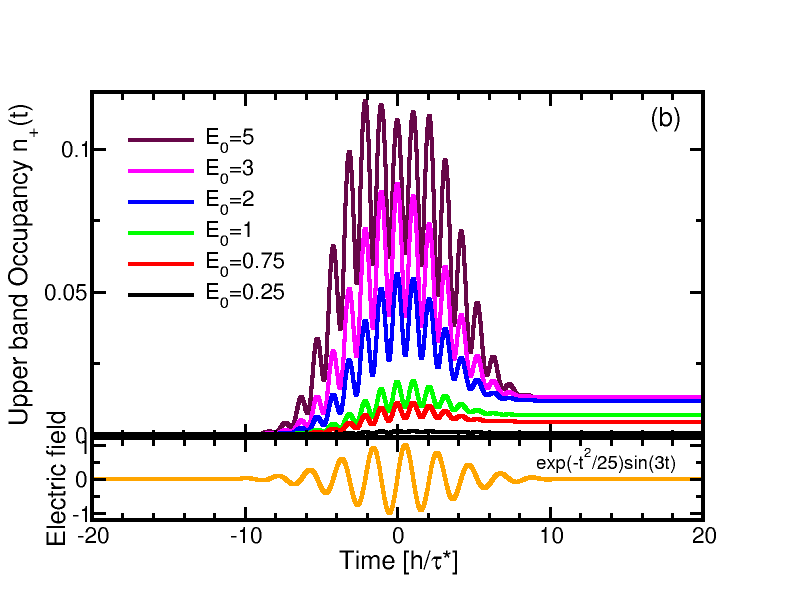}
\end{center}
\caption{(Color online) 
Excitation in the upper band as a function of time for the (a) low frequency $\omega_0=1$ and (b) high frequency $\omega_0=3$ pulsed-field cases. The labels in panel (a) and the legend in panel (b) show the amplitude of the field, which is plotted for unit amplitude below each panel.
\label{fig: pulse_excite}}
\end{figure}

The electric field is chosen to lie along the diagonal direction and each spatial component satisfies $E(t)=E_0 \sin(\omega_0t) \theta(t)$ for the monochromatic ac field and $E(t)=E_0 \sin(\omega_0t) \exp(-t^2/25)$ for the pulsed field. In Fig.~\ref{fig: ac_excite} (a) and (b), we examine the excitation process in the ac field with $\omega_0=1$ and $\omega_0=3$, respectively, for a CDW with a gap satisfying $U=1.5$. Because an ac driving field produces a periodic response, we must determine the average occupancy of the ``steady state'' which is shown by the dashed lines in Fig.~\ref{fig: ac_excite}.  Both systems approach this ``steady state'' relatively quickly, but in the lower frequency case (panel a), the final density in the upper band does not have a monotonic dependence on the driving amplitude. 

In Fig.~\ref{fig: pulse_excite} (a) and (b), we show similar plots, but now for the pulsed case. Here we see quite different behavior.  First off, a true steady state occurs, because at long times there is no field and the system dephases into a steady population in each band (transfer between bands can only occur when a field is on). Second, in the lower frequency case (panel a), one can see the excitation is dominated by the regions where the field amplitude is the largest, but the full excitation occurs over an extended period of time, and certainly is not instantaneous. Third, the excitation is fairly monotonic in time, implying it is dominated by excitation processes and there is limited de-excitation. Note the step-like excitation for low-amplitude fields, which follow precisely the Landau-Zener picture of tunneling enhanced when the instantaneous magnitude of the field is large. In contrast, the higher frequency case in panel (b), shows very dramatic de-excitation processes, and the final excitation requires one to examine the full time dependence of the system.  It cannot be described just by the regions where the field amplitude is maximal. Instead, the quantum excitation is primarily determined by the lower field amplitudes near the start of the pulse, and the rest of the evolution corresponds to nearly equal excitation followed by de-excitation. 

\begin{figure}[!ht]
\begin{center}
\includegraphics[scale=0.3,clip=on]{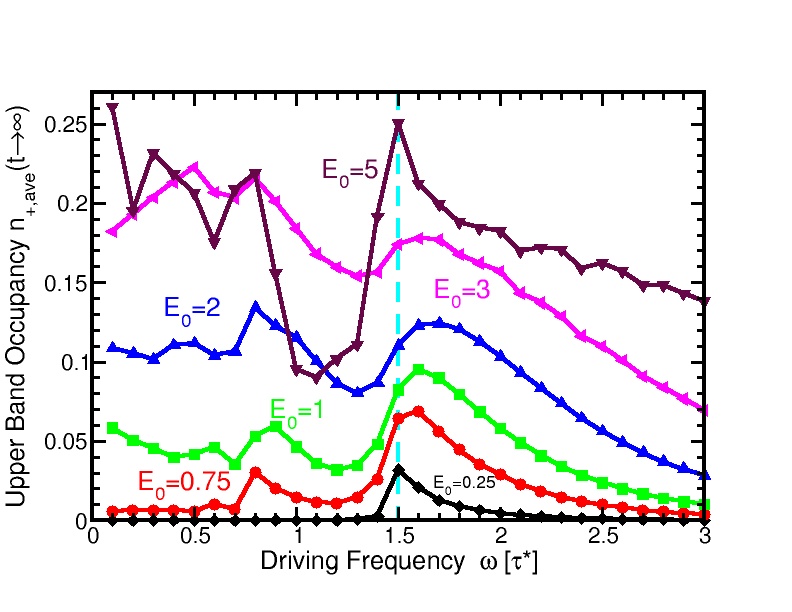}
\end{center}
\caption{(Color online) 
Upper band occupancy spectra for fixed amplitude and varying driving frequency with an ac field drive and the light blue dashed line showing where the equilibrium gap is.
\label{fig: ac_spectra}}
\end{figure}

\begin{figure}[!ht]
\begin{center}
\includegraphics[scale=0.3,clip=on]{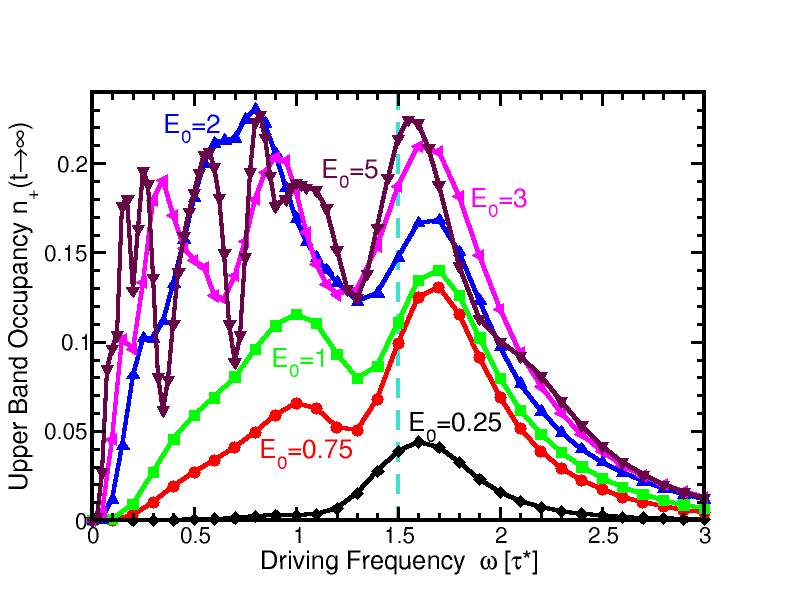}
\end{center}
\caption{(Color online) 
Excited state spectroscopy for the pulsed field with the dashed blue line showing the value of the equilibrium gap.  Here, the behavior is similar to Fig.~\ref{fig: ac_spectra}, except the frequencies are less well defined because of the finite pulse width, which becomes particularly important in the low-frequency regime, and suppresses the signal there.  The amplitude of each pulse is given by the color-coded labels near each respective curve.
\label{fig: pulse_spectra}}
\end{figure}

We now examine spectroscopy for the long-time excitation curves as a function of the driving frequency for an ac drive (Fig.~\ref{fig: ac_spectra}) and for the pulsed field (Fig.~\ref{fig: pulse_spectra}).  In both cases, when the amplitude is small, the Planck-Einstein relation holds and no excitation occurs until the frequency becomes approximately equal to the gap (there is a small spread due to the finite spread of the Fourier transform of the applied electric field, especially for the pulsed case). As the amplitude is increased, we see the expected nonlinear effect of a peak forming at a frequency about one half of the gap size. But as the amplitude is increased further, the excitation spectra become quite flat in frequency, indicating the crossover to an amplitude driven excitation, although the curvature of the bandstructure never allows for a fully flat curve.  Finally, as the amplitude is made even larger, we see interesting quantum oscillations develop in the spectra, which become more complex for larger amplitude driving. This behavior resembles the dynamical behavior shown by Shirley for two-level systems~\cite{shirley}, and the appearance could be simply a result of the renormalization of the energies of multiphoton processes and a narrowing of the peaks with increasing amplitude.  But it might also signal the onset of a new quantum regime related to localized Wannier-Stark physics that is governed primarily by the field amplitude and not the driving frequency.  

\paragraph{Summary} 
In this work, we have shown the crossover for how a quantum system is initially excited by the Planck-Einstein quanta, but then nonlinear effects change the behavior first into excitations arising from the nonlinear combination of photons to create high energy excitations until the excitation is dominated by the amplitude of the driving field and shows limited frequency dependence.  At the largest amplitudes, additional quantum oscillations occur, whose origins are likely due to novel high-field quantum effects. This work opens up a new field in examining the details behind quantum excitation in strong fields that goes beyond the common belief that tunneling phenomena are dominated by the region in time where the field amplitude is the largest.  We clearly see more complex and rich phenomena arises in this limit.

\paragraph{Acknowledgments}
The development of the parallel computer algorithms for the quantum excitation calculations was supported by the National Science Foundation under Grant No. OCI-0904597. The data analysis and application to experiment was supported by the Department of Energy, Office of Basic Energy Research under Grants No. DE-FG02-08ER46542 (Georgetown), DE-AC02-76SF00515 (Stanford/SLAC), and DE-FG02-08ER46540 and DE-SC0007091 (for the collaboration). High performance computer resources utilized the National Energy Research Scientific Computing Center supported by the Department of Energy, Office of Science, under Contract No. DE- AC02-05CH11231. J.K.F. was also supported by the McDevitt bequest at Georgetown. The research was completed during a visit to KITP and was supported in part by the National Science Foundation under Grant No. NSF PHY11-25915. We also acknowledge useful conversations with C. Kollath, P. Littlewood, and D. Scalapino.


\begin{thebibliography}{99.}
\bibitem{planck}
M. Planck,  Ann. der Phys. (Leipzig) {\bf 4}, 553 (1901).
\bibitem{einstein}
A. Einstein,  Ann. der Phys. (Leipzig)  {\bf 17}, 132 (1905).
\bibitem{landau}
L. D.  Landau, Phys. Z. Sowjetunion {\bf 2}, 46 (1932). 
\bibitem{zener}
C.  Zener, Proc. R. Soc.( London), Ser. A {\bf 137}, 696 (1932).
\bibitem{kubo}
R.  Kubo, J. Phys. Soc. Japan {\bf 12}, 570 (1957).
\bibitem{greenwood}
D. A. Greenwood,  Proc. Phys. Soc. (London) {\bf 71}, 585 (1958).
\bibitem{keldysh}
 L. V.  Keldysh,  Zh. Eksp. Teor. Fiz. {\bf 47}, 1945 (1964); [Eng. transl. Sov. Phys. JETP {\bf 20}, 1307 (1965)].
\bibitem{experiment} G.  Herink, D. R. Solli, M. Gulde, and C.  Ropers,  Nature {\bf 483}, 190 (2012).
\bibitem{peierls}
R. E. Peierls, Z. Phys. {\bf  80}, 763 (1933).
\bibitem{cdw_long} W. S. Shen, T. P. Devereaux, and J. K. Freericks, preprint arxiv:1308.6060 (2013).
\bibitem{shirley}
J. H. Shirley, Phys. Rev. {\bf 138}, B 979 (1965).

\end{thebibliography}
\end{document}